\documentstyle[seceq,supplement,epsf]{ptptex}

\newcommand{\sol}{M_{\odot}}

\newcommand{\rscript}[1]{\mbox{\scriptsize #1}}
\newcommand{\rhoh}{\rho_{_H}}
\newcommand{\rhob}{\rho_{_B}}
\newcommand{\rhon}{\rho_{_N}}

\newcommand{\gammaT}{\widetilde{\gamma}}
\newcommand{\GammaT}{\widetilde{\Gamma}}
\newcommand{\phiH}{\widehat{\phi}}

\newcommand{\KT}{\widetilde{K}}
\newcommand{\KH}{\widehat{K}}
\newcommand{\JT}{\widetilde{J}}
\newcommand{\DT}{\widetilde{D}}
\newcommand{\RT}{\widetilde{R}}
\newcommand{\LaplaceT}{\widetilde{\triangle}}
\newcommand{\auz}{\alpha u^0}
\newcommand{\sgamma}{\sqrt{\gamma}}

\newcommand{\Dtv}[2]{\partial_t #2 #1^{\ell} \partial_\ell #2}
\newcommand{\TT}[1]{#1^{\mbox{\scriptsize TT}}}
\newcommand{\htt}{\TT{h}}

\title{3D General Relativistic Simulations of Coalescing \\
  Binary Neutron Stars}

\author{%
Ken-ichi {\sc Oohara}
and
Takashi {\sc Nakamura}$^*$
}

\inst{%
Department of Physics, Niigata University, Niigata 950--2181, Japan \\
$^*$ Yukawa Institute for Theoretical Physics, Kyoto University, Kyoto
606-8502, Japan}

\abst{%
We develop a 3 dimensional computer code to study a coalescing neutron 
star binary. The code can currently follow the evolution up to two
stars begin to merge from two spherical stars of mass $1\sol$ and
radius 8.9km with separation 35.4km. As for coordinate conditions, we
use conformal slicing and pseudo-minimal distortion conditions.
The evolution equations for the
metric is integrated using the CIP method while the van Leer's scheme
is used to integrate the equations for the matter. We present a few
results of our simulations including gravitational radiation.}

\begin{document}

\maketitle

\section{Introduction}

We report on the present status of our computer code development to
study the fully general relativistic evolution of spacetimes and
matter. We solve the complete set of the Einstein equations in the
absence of any symmetries. The main motivation of this project is
studying promising sources of strong gravitational waves, such as
coalescing neutron star or black hole binaries. The first detection by 
an interferometric detector of gravitational waves such as LIGO, VIRGO,
GEO600 and TAMA may be the waves from a black hole binary (see the
contributions of K.S.~Thorne, S.~Kawamura and B.~Schutz in this
volume), but the astrophysical importance of a coalescing neutron star
binary is not smaller than a black hole binary. A detailed analysis 
of the detected waves near the merger of two stars, for example, will
give information on the size of a neutron star and then on the
equation of state of high density matter. On problems other than the
gravitational wave physics, study on process of coalescence of a
neutron star binary is helpful for models of a source of a gamma ray
burst (see the contribution of P.I.~Meszaros in this volume).

A compact star binary in the earliest stage is almost stationary but
gravitational radiation reaction makes it evolve in the adiabatic
manner. When the separation between stars is large enough, the stars
can be regarded as point particles. As the separation is getting
small, however, the size effect of the stars and the relativistic
effects become important.
Thus the complete set of the Einstein equations and the general
relativistic hydrodynamics equations for a neutron star binary must be 
solved. In addition, the process of coalescence of a
binary is not axisymmetric. Thus a three dimensional (in space) code
is necessary and it requires a great power of computers. In 1994, we
became able to use a supercomputer with memories of tens of GBytes and 
a peak speed of nearly a hundred GFLOPS. With this computer, we can
manage a $200^3$ grid. Then we started to make a 3D,
fully general relativistic code for a coalescing neutron star binary.
In this paper, we show the present status of our code development.
The present code does not include a lot of realistic physics processes
such as neutrino emission and a realistic equation of state near and
beyond the nuclear density. So it is a milestone of 3D numerical
relativity. In \S 2, we describe the mathematical framework of the
calculation, including constraint equations, evolution equations of
the metric, relativistic hydrodynamics equations, coordinate
conditions and evaluation of the gravitational waves. In \S 3,
numerical methods we use for elliptic equations and evolution
equations are shown. In \S 4, we present numerical results of
simulations of coalescing neutron star binaries obtained using the
present code. In \S 5, we summarize and outline future work planned
for this code.

\section{The Mathematical Framework}

\subsection{Constraint equations}

In the (3+1) (or ADM) formalism of general relativity, the spacetime
metric is written as
\begin{equation}
  \label{eq:fullmetric}
  ds^2 = - \alpha^2 dt^2 + \gamma_{ij} ( dx^i + \beta^i dt )
  ( dx^j + \beta^j dt) ,
\end{equation}
where $\alpha$, $\beta^i$ and $\gamma_{ij}$ are the lapse function, the 
shift vector and the intrinsic metric of three-metric. The Einstein
equation reduces to the four constraint equations
\begin{equation}
  R + K^2 - K_{ij} K^{ij} = 16 \pi \rhoh ,
  \label{eq:hconst}
\end{equation}
\begin{equation}
  D_j \left( K^j{}_i - \delta^j{}_i K \right) = 8 \pi J_i
  \label{eq:mconst}
\end{equation}
and 12 evolution equations
\begin{equation}
  \label{eq:evolgij}
  \partial_t \gamma_{ij} = - 2\alpha K_{ij} + D_i \beta_j + D_j
  \beta_i ,
\end{equation}
\begin{eqnarray}
  \partial_t K_{ij} & = & \alpha \left[ R_{ij} - 8\pi \left\{ S_{ij} +
      \mbox{$\frac{1}{2}$} \gamma_{ij} \left( \rhoh - S^{\ell}{}_{\ell} \right) 
      \right\} \right] - D_i D_j \alpha \nonumber \\
    & & + \alpha \left( K K_{ij} - 2 K_{i\ell} K^{\ell}{}_j \right)
    + K_{\ell i} D_j \beta^\ell + K_{\ell j} D_i \beta^\ell +
    \beta^\ell D_\ell K_{ij} ,
  \label{eq:evolkij}
\end{eqnarray}
where $R_{ij}$ is the Ricci tensor, $K_{ij}$ the extrinsic curvature
tensor, $K = K^{\ell}{}_{\ell}$, and $D_i$ the covariant derivative
associated with $\gamma_{ij}$. The quantities $\rhoh$, $J_i$ and
$S_{ij}$ are the energy density, the momentum density and the stress
tensor, respectively, measured by the observer moving along the line
normal to the spacelike hypersurface of $t=$constant.

In order to give an initial data, we should find a three-metric and
extrinsic curvature which satisfy the constraint equations
(\ref{eq:hconst}) and (\ref{eq:mconst}). It means that the constraint
equations are solved giving $\rhoh$ and $J_i$. Here we assume that
$K=0$ and $\gamma_{ij}$ is conformally flat at $t=0$
\begin{equation}
  \label{eq:cflat}
  \gamma_{ij} = \phi^4 \, \gammaT_{ij} ,
\end{equation}
where $\gammaT_{ij}$ is the flat space metric. Defining the conformal
transformation as
\begin{equation}
  \label{eq:conf-i}
  \begin{array}[c]{l}
    \KT_{ij}  \equiv \phi^2     K_{ij} , \qquad
    \KT_i{}^j \equiv \phi^6     K_i{}^j , \qquad
    \KT^{ij} \equiv  \phi^{10}  K^{ij} , \\
    \rhob \equiv \phi^6  \rhoh \qquad \mbox{and} \qquad
    \JT_i \equiv \phi^6  J_i ,
  \end{array}
\end{equation}
Eq.(\ref{eq:mconst}) reduces to
\begin{equation}
  \label{eq:mconst2}
  \DT_j \KT^j{}_i = 8 \pi \JT_i ,
\end{equation}
where $\DT_i$ is the covariant derivative associated with
$\gammaT_{ij}$.\cite{York79} The traceless extrinsic curvature can be
decomposed with the transverse traceless part
$\KT^{\rscript{TT}}_{ij}$ and the longitudinal traceless part
$(LW)_{ij}$ \cite{York73};
\begin{equation}
  \KT_{ij} = \KT^{\rscript{TT}}_{ij} + (LW)_{ij} ,
\end{equation}
where
\begin{equation}
  (LW)_{ij} = \DT_i W_j + \DT_j W_i
  - \frac{2}{3} \, \gammaT_{ij} \DT^{\ell} W_{\ell} .
\end{equation}
Assuming $\KT^{\rscript{TT}}_{ij} = 0$, Eq.(\ref{eq:mconst2}) becomes
\begin{equation}
  \label{eq:mconst3}
  \LaplaceT W_i + \frac{1}{3} \DT_i \DT^j W_j =
  8 \pi \JT_i ,
\end{equation}
where $\widetilde{\triangle} \equiv \DT^i \DT_i$.  Equation
(\ref{eq:mconst3}) is the coupled elliptic equation but can be
reduced to four decoupled Poisson equations:
\begin{equation}
  \label{eq:spot2}
  \triangle \chi = 6 \pi \partial_i \JT_i ,
\end{equation}
\begin{equation}
  \label{eq:vpot2}
  \triangle W_i = 8 \pi \JT_i - \frac{1}{3} \partial_i \chi ,
\end{equation}
where $\chi = \partial_i W_i$.

The Hamiltonian constraint equation (\ref{eq:hconst}) reduces to
the nonlinear Poisson equation
\begin{equation}
  \label{eq:hconst2}
  \LaplaceT \phi = -2 \pi \phi^{-1} \rhob - \frac{1}{8} \phi^{-7}
  \KT_{ij} \KT^{ij}. 
\end{equation}

\subsection{Evolution equations of the metric}

Defining the following variables
\begin{equation}
  \label{eq:conf-e1}
   \phi = \left( \mbox{det} (\gamma_{ij}) \right)^{\frac{1}{12}},
   \qquad
   \gammaT_{ij} = \phi^{-4} \gamma_{ij} \  , \qquad
   \gammaT^{ik} \gammaT_{kj} = \delta^i{}_j ,
\end{equation}
\begin{equation}
  \label{eq:conf-e2}
   \KH_{ij} = \mbox{STF} (\phi^{-4} K_{ij}), \qquad
   \KH^i{}_j = \gammaT^{ik} \KH_{kj} \  , \qquad
   \KH^{ij} = \gammaT^{jk} \KH^i{}_k ,
\end{equation}
where
\begin{equation}
  \label{eq:STF}
    \mbox{STF} (A_{ij}) \equiv \frac{1}{2} \left( A_{ij} + A_{ji} -
    \frac{2}{3} \gammaT_{ij} \gammaT^{k\ell} A_{k\ell} \right) ,
\end{equation}
the evolution equation (\ref{eq:evolkij}) reduces to
\begin{eqnarray}
  \Dtv{-\beta}{\KH_{ij}} & = &
   \phi^{-4} \left[ \mbox{STF} \left\{ \alpha \left( R_{ij} - 8 \pi
        S_{ij} \right) - D_i D_j \alpha  \right\} \right] \nonumber \\
  & + & \alpha \left( K \KH_{ij} - 2 \KH_{i\ell} \KH^\ell{}_j \right)
    \label{eq:evolkHij} \\
  & + & \KH_{i\ell} \partial_j \beta^\ell + \KH_{j\ell} \partial_i \beta^\ell
  - \frac{2}{3} \KH_{ij} \partial_\ell \beta^\ell , \nonumber \\
  \Dtv{-\beta}{K} & = & \alpha \! \left[ \KH_{ij} \KH^{ij} \! + \!
    \frac{1}{3} K^2 
    \! + \! 4\pi \! \left( \rhoh \! + \! S^i{}_i \right) \right]
  \! - \! D^i D_i \alpha , \label{eq:evoltrK}
\end{eqnarray}
and Eq.(\ref{eq:evolgij}) to
\begin{equation}
  \label{eq:evolgTij}
  \partial_t \gammaT_{ij} = - 2 \left[ \alpha \KH_{ij}
  - \mbox{STF} \left\{ \DT_i \left( \phi^{-4} \beta_j \right)
  \right\} \right]
  \equiv A_{ij}
\end{equation}
or
\begin{equation}
    \Dtv{-\beta}{\gammaT_{ij}}  = 
  - 2 \! \left[ \alpha \KH_{ij} 
  - \mbox{STF} \left( \gammaT_{i\ell} \partial_j \beta^\ell
  \right) \right] \label{eq:evolgTij2} .
\end{equation}

The conformal factor $\phi$ obeys
\begin{equation}
  \label{eq:evolphi}
  \Dtv{- \beta}{\phi} = - \frac{\phi}{6}
  \left( \alpha K - \partial_\ell \beta^\ell \right)
\end{equation}
or
\begin{equation}
  \label{eq:hconst-e}
    \LaplaceT \phi = - \frac{\phi^5}{8} \left( 16 \pi \rhoh + K_{ij} K^{ij} -
    K^2 - \phi^{-4} \RT \right) .
\end{equation}
Equation (\ref{eq:evolphi}) follows from the trace of
Eq.(\ref{eq:evolgij}), while Eq.(\ref{eq:hconst-e}) from the
Hamiltonian constraint (\ref{eq:hconst}).

\subsection{Relativistic hydrodynamics equations}

We assume the perfect fluid stress-energy tensor, which is given by
\begin{equation}
  \label{eq:enmom}
  T_{\mu \nu} = ( \rho + \rho \varepsilon + p ) u_{\mu} u_{\nu}
  + p g_{\mu \nu},
\end{equation}
where $\rho$, $\varepsilon$ and $p$ are the proper mass density, the
specific internal energy and the pressure, respectively, and $u_\mu$
is the four-velocity of the fluid. The energy density $\rhoh$, the
momentum density $J_i$ and the stress tensor $S_{ij}$ of the matter
measured by the normal line observer are, respectively, given by
\begin{equation}
  \label{eq:qhydro}
  \rhoh \equiv n^\mu n^\nu T_{\mu \nu}, \ \
  J_i \equiv - h_i{}^\mu n^\nu T_{\mu \nu} \ \ \mbox{and} \ \ 
  S_{ij} \equiv h_i{}^\mu h_j{}^\nu T_{\mu \nu},
\end{equation}
where $n_\mu$ is the unit timelike four-vector normal to the spacelike 
hypersurface and $h_{\mu \nu}$ is the projection tensor into the
hypersurface defined by
\begin{equation}
  h_{\mu \nu} = g_{\mu \nu} + n_\mu n_\nu .
\end{equation}
The relativistic hydrodynamics equations are obtained from the
conservation of baryon number, $\nabla_\mu ( \rho u^\mu )$, and the
energy-momentum conservation law, $\nabla_\nu T_\mu{}^\nu{}$. In order
to obtain equations similar to the Newtonian hydrodynamics equations,
we define $\rhon$ and $u_i^N$ as
\begin{equation}
  \label{eq:rhonun}
  \rhon \equiv \sgamma \auz \rho \ \ \  \mbox{and} \ \ \  
  u_i^N = \frac{J_i}{\auz \rho},
\end{equation}
respectively, where $\gamma = \mbox{det} (\gamma_{ij})$. Then the
equation for the conservation of baryon number takes the form
\begin{equation}
  \label{eq:hydrob}
  \partial_t \rhon +   \partial_\ell \left( \rhon V^\ell \right) = 0 ,
\end{equation}
where
\begin{equation}
  V^\ell = \frac{u^\ell}{u^0} = \frac{\alpha J^\ell}{p + \rhoh} -
  \beta^\ell .
\end{equation}
The equation for momentum conservation is
{\arraycolsep = 2pt
\begin{eqnarray}
  \partial_t (\rhon u_i^N) +
  \partial_\ell \left( \rhon u_i^N V^\ell \right)
  & = & - \sgamma \alpha \partial_i p - \sgamma ( p + \rhoh )
  \partial_i \alpha \nonumber \\
  & & \frac{\sgamma \alpha J^k J^\ell}{2(p + \rhoh)} \partial_i
  \gamma_{k \ell} + \sgamma J_\ell \partial_i \beta^\ell .
  \label{eq:hydrom}
\end{eqnarray}
}%
The equation for internal energy conservation becomes
\begin{equation}
  \label{eq:hydroe}
  \partial_t (\rhon \varepsilon) +
  \partial_\ell \left( \rhon \varepsilon V^\ell \right) =
  - p \partial_\nu \left( \sgamma \alpha u^\nu \right).
\end{equation}
To complete hydrodynamics equations, we need an equation of state,
\begin{equation}
  \label{eq:eos}
  p = p(\varepsilon, \rho).
\end{equation}

The right-hand side of Eq.(\ref{eq:hydroe}) includes the time
derivative. For a polytropic equation of state, $p = (\Gamma - 1) \rho 
\varepsilon$, however, the equation reduces to
\begin{equation}
  \label{eq:hydroe2}
    \partial_t (\rhon \varepsilon_{_N}) +
  \partial_\ell \left( \rhon \varepsilon_{_N} V^\ell \right) =
  - p_{_N} \partial_\ell V^\ell ,
\end{equation}
where
\begin{eqnarray}
  \label{eq:defen}
  \varepsilon_{_N} & = & \left( \sgamma \auz \right)^{\Gamma -1} \,
  \varepsilon , \\
  \label{eq:defpn}
  p_{_N} & = & (\Gamma - 1) \rhon \varepsilon_{_N} = \left( \sgamma \auz
   \right)^\Gamma p .
\end{eqnarray}

\subsection{Coordinate conditions and gravitational waves}

The choice of a shift vector $\beta^i$ and a lapse function $\alpha$
is important because the stability of the code
largely depends on it and because it is intimately related to the
extraction of physically relevant information, including gravitational
radiation, in numerical relativity.

Since the right-hand side of Eq.(\ref{eq:evolgTij}), $A_{ij}$ is
trace-free, the determinant of $\gammaT_{ij}$ is preserved in time
and the condition
\begin{equation}
  \label{eq:mind}
  D^j A_{ij} = 0
\end{equation}
produces the minimal distortion shift vector.\cite{YS78} It must
be a good choice of the spatial coordinate, but it is too complicated
to be solved. Instead we replace the covariant derivative by the
partial derivative
\begin{equation}
  \label{eq:pmind}
  \partial_i A_{ij} = 0 ,
\end{equation}
which yields an elliptic equation for the shift vector $\beta^i$:
\begin{equation}
  \label{eq:beta}
\nabla^2 \beta^i + \frac{1}{3} \partial_i \partial_\ell
  \beta^\ell
   =  2 \partial_j \left[ \alpha \KH_{ij} - \mbox{STF} ( h_{i\ell}
   \partial_j \beta^\ell ) - \frac{1}{2} \beta^\ell \partial_\ell
   h_{ij} \right] ,
\end{equation}
where
\begin{equation}
  \label{eq:hdef}
  h_{ij} = \gammaT_{ij} - \delta_{ij} .
\end{equation}
We call this condition as the pseudo-minimal distortion condition.

As for slicing condition, we choose the conformal slicing,\cite{SN92}
where the lapse function $\alpha$ is given by
\begin{equation}
  \label{eq:confalpha}
  \alpha = \exp \left[ -2 \left( \phiH +
      \frac{\phiH^3}{3} + \frac{\phiH^5}{5} \right)
  \right] 
\end{equation}
with $\phiH = \phi - 1$.
Since in this slicing, 
\begin{equation}
  \alpha \approx \frac{1 - \phiH}{1 + \phiH} \rightarrow
  \frac{1 - \frac{M}{2r}}{1 + \frac{M}{2r}}
\end{equation}
for large $r$, the space outside the central matter  quickly
approaches the Schwarzschild metric in the isotropic coordinates.
Thus $h_{ij}$ defined by Eq.(\ref{eq:hdef}) can be considered as
gravitational wave parts for large $r$ and the total energy of the
gravitational waves is given by
\begin{equation}
  \label{eq:gwtotal}
  E_{\rscript{GW}} = \int dt \int_{r \rightarrow \infty}
    d \Omega \  r^2 \, \rho_{_{\rscript{GW}}} ,
\end{equation}
\begin{equation}
  \label{eq:gwdens}
  \rho_{_{\rscript{GW}}} = \frac{1}{32\pi} A_{ij} A_{ij} ,
\end{equation}
since Eq.(\ref{eq:pmind}) means that $A_{ij}$ is the
transverse-traceless part of the time derivative of $\gammaT_{ij}$ for 
large $r$.

\section{Numerical Methods}

\subsection{Coordinate system}

As for the spatial coordinates, we use a Cartesian coordinate system
$(x, y, z)$. We solve partial differential equations (PDEs) using a
finite-difference scheme, where a finite grid is introduced and the
PDEs are replaced by finite-difference approximations. It may be
convenient to use a spherical coordinate system $(r, \theta, \phi)$
with regard to saving of computer memory and an outer boundary
condition at large distance from a compact source. But the use of
spherical coordinates leads to a vexing problem of coordinate
singularities at the origin and along the polar axis.

\subsection{Elliptic equations} \label{sec:ellip}

The constraint equations reduce four Poisson equations
(\ref{eq:spot2}) and (\ref{eq:vpot2}), and a nonlinear Poisson equation 
(\ref{eq:hconst2}). These equations are solved once for initial data.
The shift vector is given by the coupled 
elliptic equation (\ref{eq:beta}), which is solved at every time step.
The conformal factor $\phi$ is, in turn, determined either by
the hyperbolic equation (\ref{eq:evolphi}) or by the elliptic equation
(Hamiltonian constraint) (\ref{eq:hconst-e}). However, since it is
known that the conformal slicing yields instability in $\phi$ and
$\alpha$,\cite{SN95} we solve the Hamiltonian constraint
(\ref{eq:hconst-e}) at every time step to prevent the onset of the
instability caused by numerical errors from occurring quickly.
Thus the solution of the elliptic equations consumes the greatest part
of the CPU time.

A non-linear equation such as
\begin{equation}
  \label{eq:NLPois}
  \triangle \phi(t) = -4 \pi S(\phi(t)) ,
\end{equation}
where $S(\phi)$ is a non-linear function of $\phi$, must be solved by
an iterative method:
\begin{equation}
  \label{eq:NLiteration}
  \phi^{(I + 1)} = \triangle^{-1} S\left(\phi^{(I)}\right)
  \quad \mbox{for $I = 0, 1, 2, \cdots$}.
\end{equation}
To solve Eq.(\ref{eq:hconst2}) for initial data, we should made the
iteration. For Eq.(\ref{eq:hconst-e}), however, we don't made the
iteration to reduce the required CPU time. Instead, the right-hand side
$S(\phi(t))$ is calculated using an approximate value of $\phi(t)$
evaluated by extrapolation from the values of $\phi$ at the previous two
time steps $t - \Delta t$ and $t - 2 \Delta t$. We examined the
difference of the solution using the extrapolated source from the solution
obtained by the iterative method, and found that it is small enough if 
$\Delta t$ is as small as we use for our actual calculations.

We use the extrapolated value of $\beta^i$ in calculating the
right-hand side of Eq.(\ref{eq:beta}), too. Note that terms including
$\beta^i$ vanish and we don't need the extrapolated value at $t=0$,
since $h_{ij} = 0$. Thus Eq.(\ref{eq:beta}) reduces four Poisson
equations (\ref{eq:spot2}) and (\ref{eq:vpot2}), replacing $W_i$
to $\beta^i$.

Now all the elliptic equations reduce Poisson equations, which are
solved using a pre-conditioned conjugate gradient method.\cite{ONS97}
The boundary condition of $\phi$ giving by Eq.(\ref{eq:NLPois}) is that
\begin{equation}
  \label{eq:boundphi1}
  \phi = \frac{M}{r} + \frac{d_x x^k}{r^3} +
  O\left(\frac{1}{r^3}\right) ,
\end{equation}
for large $r$, where
\begin{equation}
  M = \int S \, d^3 x, \qquad
  d_k  =  \int S x^k \, d^3 x \label{eq:poisMd} .
\end{equation}
The boundary condition of $\chi$ and $W_i$ (or $\beta^i$) giving by
Eqs.(\ref{eq:spot2}) and (\ref{eq:vpot2}) is that\cite{ONS97}
\begin{eqnarray}
  \label{eq:spot2b}
  \chi & = & \frac{P_i x^i}{2 r^3} - \frac{3 M_{ii}}{2 r^3} + \frac{9M_{ij} 
      x^i x^j}{2 r^5} + O\left(\frac{1}{r^4}\right), \\
  W_i & = & - \frac{P_k x^k x^i}{4 r^3} - \frac{7 P_i}{4 r} \nonumber
  \\
  & & - \frac{\left(7 M_{ij} - M_{ji} - M_{kk} \delta_{ij} \right)
    x^j}{4 r^3} - \frac{3 M_{jk} x^j x^k x^i}{4 r^5} 
  + O \left(\frac{1}{r^3} \right) \label{eq:vpot2b}
\end{eqnarray}
for large $r$, where
\begin{equation}
  P_i  =  \int \JT_i \, d^3 x , \qquad
  M_{ij}  =  \int \JT_i x^j \, d^3 x \label{eq:totalmom} .
\end{equation}

\subsection{Evolution of matter}

Hydrodynamics equations Eqs.(\ref{eq:hydrob}), (\ref{eq:hydrom}) and
(\ref{eq:hydroe2}) are written as
\begin{equation}
  \label{eq:hydroQ}
  \partial_t Q + \partial_\ell ( Q V^\ell ) = F ,
\end{equation}
which is split into two phases:
\begin{equation}
  \label{eq:hydroa}
  \partial_t Q + \partial_\ell ( Q V^\ell ) = 0 ,
\end{equation}
\begin{equation}
  \label{eq:hydron}
  \partial_t Q = F .
\end{equation}
Here we call Eq.(\ref{eq:hydroa}) the advection phase and
Eq.(\ref{eq:hydron}) the non-advection phase. The advection phase is
solved using van Leer's scheme\cite{vanLeer} to obtain
$Q^{(n+1)*}_{ijk}$ from $Q^n_{abc}$ and $(V^\ell)^n_{abc}$, where
$Q^n_{abc}$ is the value of $Q$ at $n$-th time step $t = t^n$ and a
spatial grid point $(x^a, y^b, z^c)$.
If the non-advection phase is solved as
\begin{equation}
  \label{eq:hydron1}
  Q^{n+1}_{ijk} = Q^{(n+1)*}_{ijk} + \Delta t \cdot  F^{n}_{ijk}
\end{equation}
with $F^{n}_{ijk}$ calculated using quantities at $t = t^n$,
it is of first-order accuracy in time. To achieve second-order
accuracy, we use (1) a two-step algorithm
\begin{eqnarray}
  \label{eq:hydron21}
  Q^{n+\frac{1}{2}}_{ijk} & = & Q^{(n+\frac{1}{2})*}_{ijk} + 
\frac{1}{2} \Delta t \cdot  F^{n}_{ijk},
\\
  \label{eq:hydron22}
  Q^{n+1}_{ijk} & = & Q^{(n+1)*}_{ijk} + \Delta t \cdot
F^{n+\frac{1}{2}}_{ijk} ,
\end{eqnarray}
where $Q^{n+\frac{1}{2}}_{ijk}$ is used only in calculating
$F^{n+\frac{1}{2}}_{ijk}$, or (2) a extrapolated source algorithm
\begin{equation}
  \label{eq:hydrone}
  Q^{n+1}_{ijk} = Q^{(n+1)*}_{ijk} + \Delta t \cdot
   \widetilde{F}^{n+\frac{1}{2}}_{ijk} ,
\end{equation}
where $\widetilde{F}^{n+\frac{1}{2}}_{ijk}$ given by
\begin{equation}
  \label{eq:fextr}
  \widetilde{F}^{n+\frac{1}{2}}_{ijk} = F^n_{ijk}
   + \frac{1}{2}(F^n_{ijk} - F^{n-1}_{ijk})
\end{equation}
is the value of $F$ at $t = t^n + \frac{1}{2} \Delta t$ evaluated by 
extrapolation from the values at $t = t^n$ and $t = t^n - \Delta t$.
Since some test calculations indicated that these algorithms give
almost the same results if $\Delta t$ is small, we will use the
extrapolated source algorithm.

Since the outer boundary is sufficiently far from the region of the
matter distribution, we can safely set $Q = 0$ at the outer boundary.

\subsection{Evolution of metric}

The evolution equations for $\KH_{ij}$ (\ref{eq:evolkHij}), $K$
(\ref{eq:evoltrK}) and $\gammaT_{ij}$ (\ref{eq:evolgTij2}) can be
written as
\begin{equation}
  \label{eq:evmetric}
  \partial_t Q + v^\ell \partial_\ell Q = F ,
\end{equation}
with $v^\ell = - \beta^\ell$.
It is solved using the CIP (Cubic-Interpolated
Pseudoparticle/Prop\-a\-ga\-tion) method.\cite{YABE97}
In the CIP method, both Eq.(\ref{eq:evmetric}) and its spatial derivatives
\begin{equation}
  \label{eq:evmetricd}
  \partial_t (\partial_a Q) + v^\ell \partial_\ell (\partial_a Q)
   = - ( \partial_i v^\ell) (\partial_\ell Q) + \partial_a F
\end{equation}
are solved. Numerical diffusion during propagation of $Q$ can be reduced,
since the time evolution of $Q$ and its derivatives are traced.
Equations (\ref{eq:evmetric}) and (\ref{eq:evmetricd}) are split into
the advection and the non-advection phases similarly to 
Eqs.(\ref{eq:hydroa}) and (\ref{eq:hydron}).
In the non-advection phase, $F$ and $\partial_a F$ at $t = t^n +
\frac{1}{2} \Delta t$ are evaluated by extrapolation from the values
at previous two time steps.

By the way, special care must be taken to calculate the Ricci tensor
appearing in the right-hand side of Eq.(\ref{eq:evolkHij}).
With conformal transformation Eq.(\ref{eq:conf-e1}), the Ricci tensor
$R_{ij}$ associated with $\gamma_{ij}$ is given by
\begin{equation}
  \label{eq:Ricci}
  R_{ij} = \widetilde{R}_{ij} + R^{(\phi)}_{ij} ,
\end{equation}
where 
\begin{eqnarray}
  R^{(\phi)}_{ij} & = & - 2 \phi^{-1} \left( \DT_j \DT_i \phi + \gammaT_{ij}
    \widetilde{\triangle} \phi \right) \nonumber \\
  & & + 2 \phi^{-2} \left( 3 (\DT_i \phi)(\DT_j \phi) - \gammaT_{ij}
   (\DT_k \phi) (\DT^k \phi) \right)
  \label{eq:Rijphi}
\end{eqnarray}
and $\widetilde{R}_{ij}$ is the Ricci tensor associated with
$\gammaT_{ij}$, which is given by
\begin{equation}
  \widetilde{R}_{ij}  =  \frac{1}{2} \left[ \gammaT^{k \ell} \left(
      h_{\ell i,jk} + h_{\ell j,ik} - h_{ij,k\ell}
    \right) + \gammaT^{k \ell}{}_{,k} \left(
      h_{\ell i,j} + h_{\ell j,i} - h_{ij,\ell}
    \right) \right] - \GammaT^{k}_{\ell i} \GammaT^{\ell}_{jk} .
  \label{eq:RTij}
\end{equation}
The second derivatives of $h_{ij}$ are replaced by finite differences
in numerical calculation but the numerical precision of terms such as
$h_{ij,k\ell}$ with $k \ne \ell$ is not so good, while the degree of
precision in calculation of $h_{ij,k\ell}$ with $k = \ell$ is the same
as that of the first derivatives.
Inaccuracies in $h_{ij,k\ell}$ will cause a
numerical instability. The `pseudo-minimal distortion condition',
Eq.(\ref{eq:pmind}), however, leads $\delta^{jk} h_{ij,k} = 0$ and
therefore
\begin{equation}
  \label{eq:RTij2}
  \gammaT^{k \ell} \left( h_{\ell i,jk} + h_{\ell j,ik} -
    h_{ij,k\ell} \right)
  = - h_{ij,kk} + \psi^{k \ell} \left( h_{\ell i,jk} + h_{\ell j,ik} -
    h_{ij,k\ell} \right) ,
\end{equation}
where $\psi^{ij} = \gammaT^{ij} - \delta^{ij}$. Inaccuracies in
numerical values of $\psi^{k \ell} h_{\ell i,jk}$ will not affect
integration of Eq.(\ref{eq:evolkHij}) so seriously since both $\psi^{k
  \ell}$ and $h_{\ell i,jk}$ are smaller than unity.

Boundary conditions for $\KH_{ij}$ and $\gammaT_{ij}$ are important,
because the boundary is not so far from the stars and therefore
inappropriate boundary conditions cause reflections of the outgoing
waves. We apply outgoing conditions for $\KT_{ij}$ and $\gammaT_{ij}$,
since non-wave parts decrease rapidly in our coordinate
conditions. The condition can be written as
\begin{equation}
  \label{eq:outgoing}
  Q(t, x^i) = \frac{H(\alpha t - \phi^2 r)}{r} ,
\end{equation}
where $r = \sqrt{x^2 + y^2 + z^2}$. Here $H$ satisfies an advection
equation
\begin{equation}
  \label{eq:outgoingF}
  \partial_t H + c^\ell \partial_\ell H = 0 ,
\end{equation}
where
\begin{equation}
  \label{eq:vwave}
  c^\ell = \frac{\alpha x^\ell}{r \phi^2} .
\end{equation}
It implies
\begin{equation}
  \label{eq:outgoingQ}
  \partial_t Q + c^\ell \partial_\ell Q = - \frac{\alpha}{r \phi^2} .
\end{equation}
Equation (\ref{eq:outgoingQ}) is solved along with
Eq.(\ref{eq:evmetric}) using the CIP method.

\section{Numerical Simulations of Coalescing Binary Neutron Stars}

Now we show numerical results on simulations of coalescing binary
neutron stars. As for a equation of state, we use the $\gamma = 2$
polytropic equation of state. To reduce the required memory, we
assume the symmetry with respect to the equatorial ($z = 0$) plane.
A uniform grid of size $191 \times 191 \times 96$ with $\Delta x =
\Delta y = \Delta z = 1 G \sol / c^2 = 1.5$km is used. The numerical
boundary is located at $95 G \sol / c^2$ from the
origin. In order to keep precision, we set the time step $\Delta
t$ as 0.01$G \sol/c^3 = 5 \times 10^{-8}$ seconds. Calculations are
performed on Fujitsu VPP500/80 at High
Energy Accelerator Research Organization (KEK) and Fujitsu VPP300/16R
at the National Astronomical Observatory (NAO), Japan. The required
memory is approximately 8GBytes.

\begin{figure}[tbp]
  \begin{center}
    \leavevmode
    \epsfxsize=.52\textwidth \epsfbox{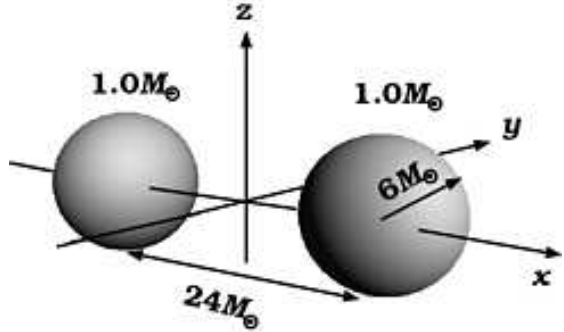}
    \caption{Initial configuration of a coalescing neutron star
      binary.}
    \label{fig:init}
  \end{center}
\end{figure}

\subsection{Initial data}

We performed calculations with two kinds of initial
data. One is the corotation (synchronized rotation) case as for
ordinary stellar binaries, in which the momentum density $\JT_i$
appearing in Eq.(\ref{eq:mconst2})  of each star is
given by
\begin{equation}
  \label{eq:vcorot}
  \JT_i = \rhob \epsilon_{ijk} \Omega^j x^k ,
\end{equation}
where $\epsilon_{ijk}$ is the complete antisymmetric symbol and the
angular velocity $\Omega^j$ is constant. The other is the
irrotational (zero-circulation) case as for gravitational radiation
driven compact stellar binaries, in which $\JT_i$ is given by
\begin{equation}
  \label{eq:virrot}
  \JT_i = \rhob \epsilon_{ijk} \Omega^j x^k_c , 
\end{equation}
where $x^k_c$ is the position of each star and is constant for an
individual star. In both cases, we assume that the binary
consists of identical spherical stars of rest mass $1.0 \sol$, radius
8.9km located at $(x_c, y_c, z_c) = (\pm 17.7\mbox{km}, 0, 0)$. The
orbital angular momentum $\Omega$ points to the $z$-direction.

\begin{table}[tbp]
  \begin{center}
    \caption{Parameters of initial data. The angular momentum
      parameter $q$ is $J_t/(G M_{\rscript{tot}}^2/c)$.}
    \label{tab:initparm}
    \begin{tabular}{|l|c|c|}
      \hline
       & Corotation & Irrotation \\ \hline
      Orbital angular velocity $\Omega$ [sec$^{-1}$] & $2.4 \times
      10^3$ & $2.4 \times 10^3$ \\
      \hline
      Angular momentum $J_t$ [$G \sol^2/c$] & 3.9 & 3.5 \\ \hline
      Collision velocity $v_x$ [$c$] & 0.074 & 0.071 \\ \hline
      Total gravitational mass $M_{\rscript{tot}}$ [$\sol$] & 1.96
      & 1.95 \\ \hline
      Angular momentum parameter $q$ & 1.03 & 0.93 \\ \hline
  \end{tabular}
  \end{center}
\end{table}

\subsection{Corotation case}

\begin{figure}[tbp]
  \begin{center}
    \leavevmode
    \epsfxsize=.75\textwidth \epsfbox{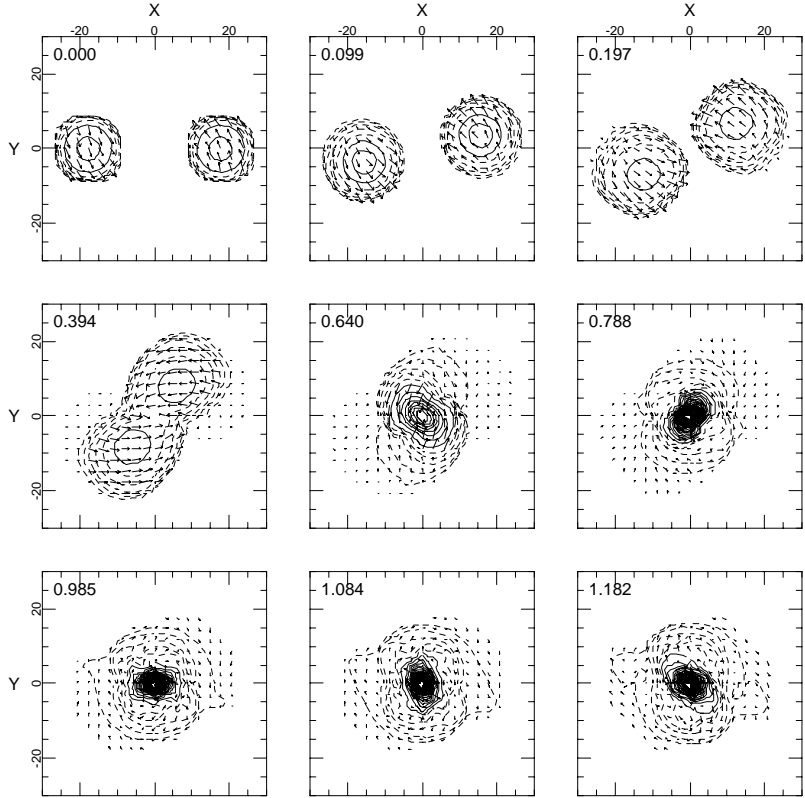}
    \caption{Density $\rhon$ and velocity $V^i$ on the $x$-$y$ plane
      for the corotation case. Time in units of milliseconds is shown.
      Solid lines are drawn from $\rho_{\rscript{min}}$ through
      $\rho_{\rscript{max}}$ in steps of $\rho_{\rscript{min}}$, where
      $\rho_{\rscript{min}} = 8 \times 10^{14}$g$\cdot$cm$^{-3}$ and
      $\rho_{\rscript{max}} = 4 \times 10^{16}$g$\cdot$cm$^{-3}$.
      Dashed lines are at 1/2, 1/4, 1/8, 1/16, 1/32 of
      $\rho_{\rscript{min}}$.}
    \label{fig:rho-cor}
  \end{center}
\end{figure}

The parameters of initial data for the corotation case are shown in
the second column of
Table~\ref{tab:initparm}. It take about 10 hours on VPP500 using 64
processor units up to 26000 time steps.
Figure~\ref{fig:rho-cor} shows the evolution of the density and the
velocity on the $x$-$y$ plane. Movies for it including the evolution
on $x$-$z$ and $y$-$z$ planes can be obtained from our
Web site.\cite{WWW} At the $\sim$26000th time step, the
onset of instability on $\phi$ in the central region occurs and it
grows exponentially in a few time steps. At the time, only 3 grid
points or so are included from the center to the surface of the
star. So that numerical errors grow on account of unstable modes
resident in the conformal slicing.

\begin{figure}[tbp]
  \begin{center}
    \leavevmode
    \epsfxsize=.45\textwidth \epsfbox{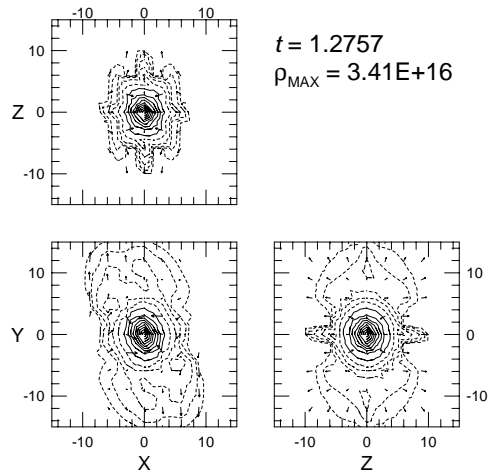}
    \caption{Density $\rhon$ and velocity $V^i$ in the central region
      on the $x$-$y$, $x$-$z$ and $z$-$y$ planes for the corotation case.
      Parameters for contour lines are the same as
      Fig.~\ref{fig:rho-cor} but
      $\rho_{\rscript{max}} = 3.41 \times 10^{16}$g$\cdot$cm$^{-3}$ and
      $\rho_{\rscript{min}} = 0.1 \rho_{\rscript{max}}$.}
    \label{fig:rhoc-cor}
  \end{center}
\end{figure}

A thick disk around the merged star is likely to be formed as shown in 
Fig.~\ref{fig:rhoc-cor}. The mass of the disk is about a few percent
of the total mass. 
In addition, there are indications of outflow along the rotation
axis. However, since the precision of the numerical calculation at
this time is not
so good, these facts must be confirmed with more precise calculation.

\begin{figure}[tbp]
  \begin{center}
    \leavevmode
    \epsfxsize=.7\textwidth \epsfbox{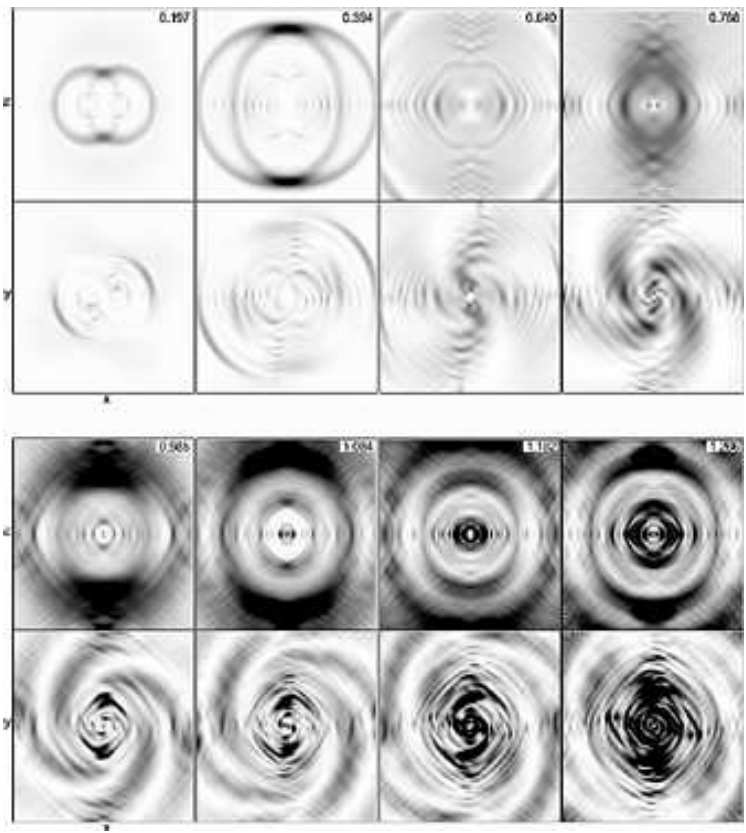}
    \caption{The propagation of the gravitational waves for the
      corotation case. The ``energy densities of gravitational waves'' $r^2
      \rho_{_{\rscript{GW}}}$ on the $x$-$y$ and $x$-$z$ planes are shown
      as gray scale figures up to the numerical boundary.
      Time in units of milliseconds is shown. The upper figure of each 
      time is on $x$-$z$ plane and the lower one is on $x$-$y$ plane.}
    \label{fig:gwxyxz-cor}
  \end{center}
\end{figure}

The propagation of the ``energy density of gravitational waves'' $r^2
\rho_{_{\rscript{GW}}}$ is shown in Fig.~\ref{fig:gwxyxz-cor}, movies 
of which are also provided on our Web site.\cite{WWW}
A spiral pattern appears on the $x$-$y$ plane, while different patterns
with peaks around $z$-axis appear on the $x$-$z$ plane.
This can be explained naively by the quadrupole wave pattern
given by
\begin{equation}
  r^2 \rho_{_{\rscript{GW}}} = \frac{r^2}{32 \pi} \left( A_{ij}
  \right)^2   \propto \cos^2 \theta + \sin^2 \theta 
  \sin^2 (2 \Omega(t - r) - 2   \varphi )/4.
\end{equation}
On the $x$-$y$ plane, where $\theta = \pi/2$, $\rho_{_{\rscript{GW}}}$ is
constant along the spiral of $r + \varphi/\Omega =$ constant, while
near $z$-axis, where $\theta \approx 0$, $\rho_{_{\rscript{GW}}}
\propto \cos^2 \theta$. 

\begin{figure}[tbp]
  \begin{minipage}[t]{.45\textwidth}
  \begin{center}
    \leavevmode
    \epsfxsize=\textwidth \epsfbox{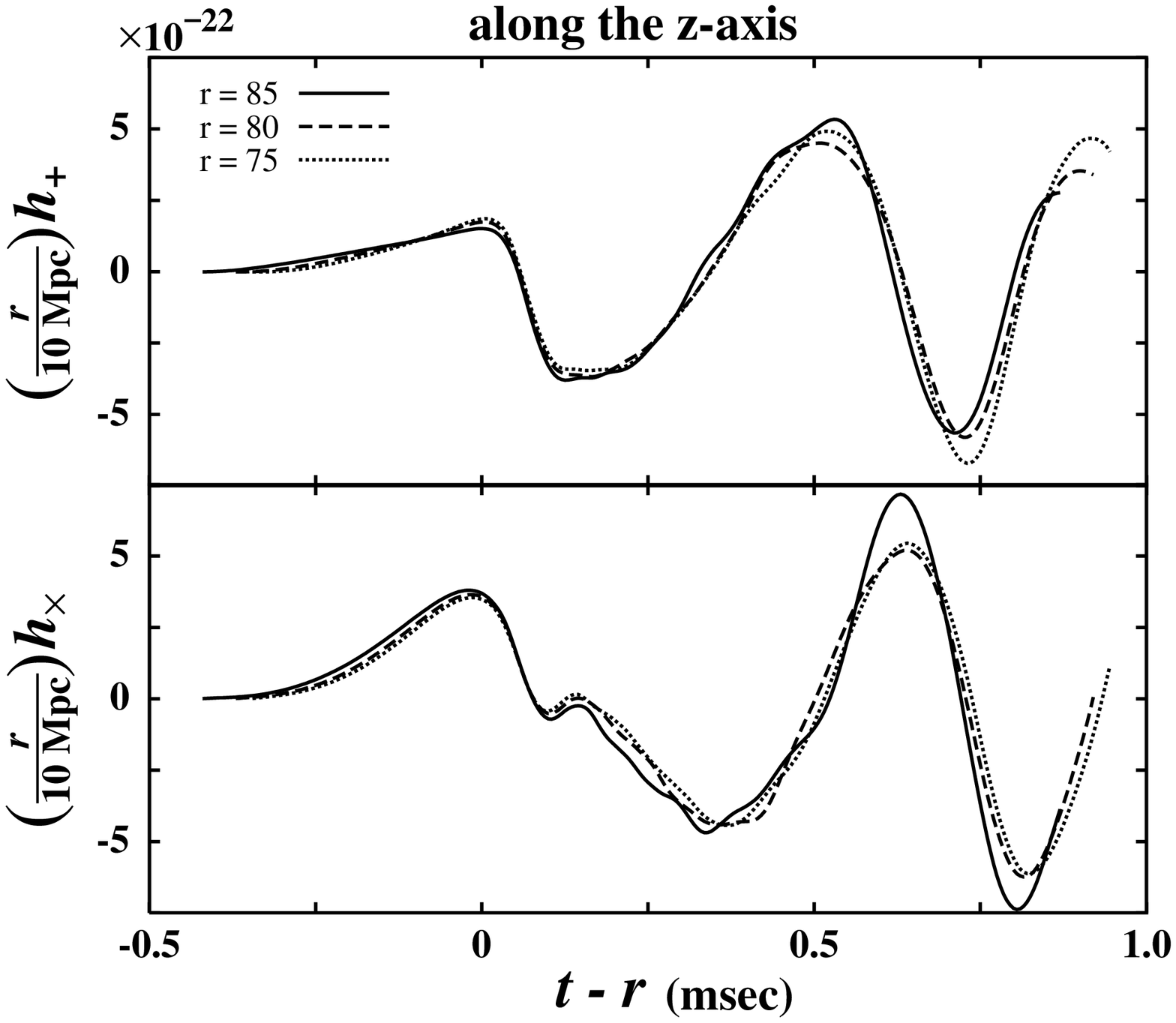}
    \caption{Waveforms $h_+$ and $h_\times$ observed on the $z$-axis
      for the corotation case. They have been monitored at $r =
      75\sol, 80 \sol$ and $85 \sol$.}
    \label{fig:wave-cor}
  \end{center}
  \end{minipage}
  \hfill
  \begin{minipage}[t]{.5\textwidth}
  \begin{center}
    \leavevmode
    \epsfxsize=\textwidth \epsfbox{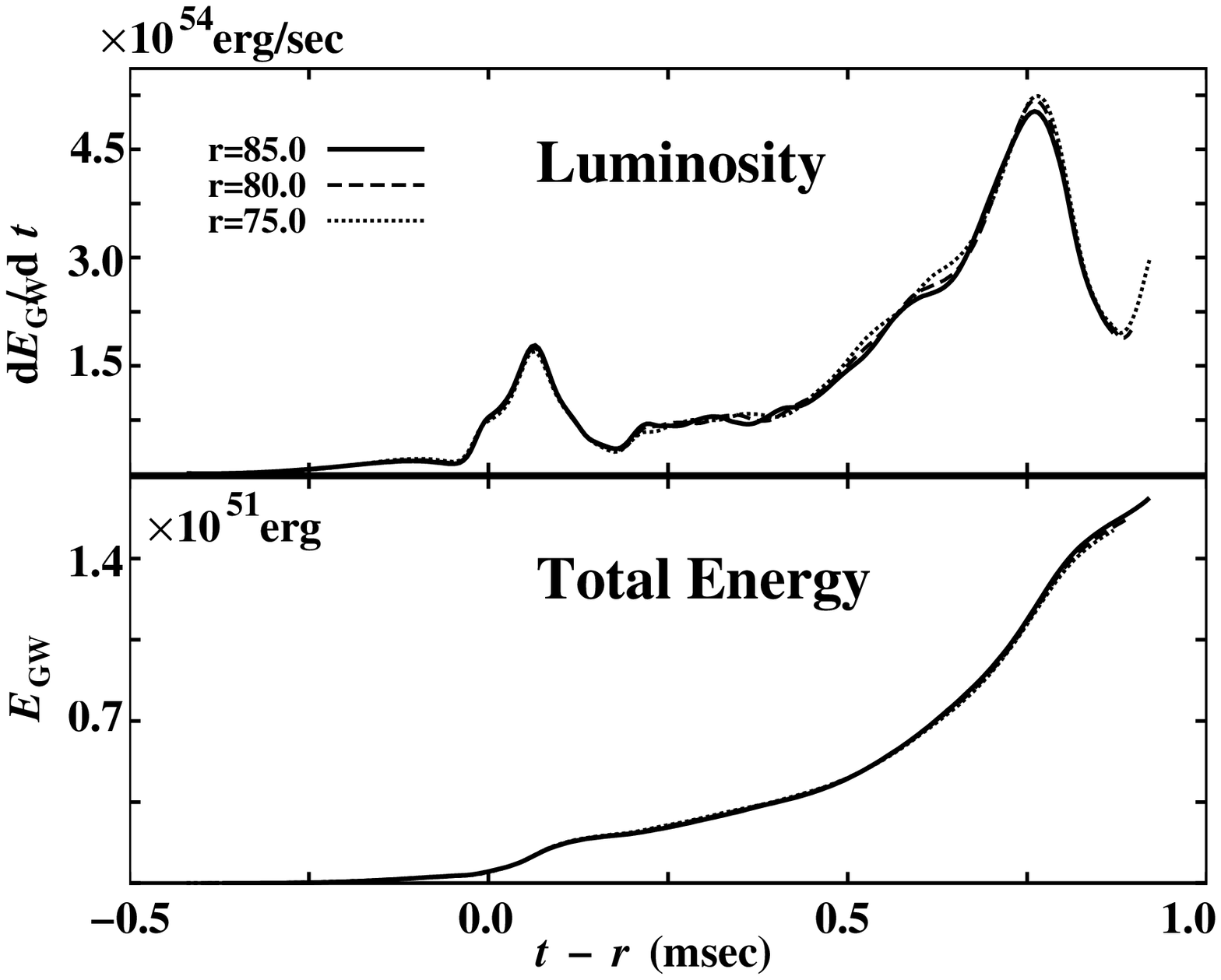}
    \caption{Luminosity (in units of erg/sec) and total energy (in
      units of erg) of the emitted gravitational waves.
      They are evaluated at $r = 75\sol, 80 \sol$ and $85 \sol$.}
    \label{fig:flux-cor}
  \end{center}
  \end{minipage}
\end{figure}

The waves pass through the numerical boundary with little
reflection. It means that the boundary condition
Eq.(\ref{eq:outgoingQ}) behaves well. However,
a small reflection on the $x$-, $y$- and $z$-axes interfere
with the outgoing waves and interference patterns appears after $t
\sim 0.8$msec.

Figures \ref{fig:wave-cor} shows the waveforms on the $z$-axis 
monitored between $75\sol$ and $85 \sol$ from the origin. The emitted
waveform falls off with $r^{-1}$ as expected. The luminosity evaluated 
between $75\sol$ and $85\sol$ as functions of $t$ and the total energy
of the gravitational waves up to $t$ are shown in
Fig.~\ref{fig:flux-cor}.
Comparing Figs.~\ref{fig:gwxyxz-cor}, \ref{fig:wave-cor} and
\ref{fig:flux-cor} with Fig.~\ref{fig:rho-cor}, the main part of the
waves begins to be emitted when the merger develops to a certain
extend, but only the waves up to the time when the merger begins
have arrived at the numerical boundary. The minor peak around $t - r = 
0$ corresponds the waves included unexpectedly in the initial data.

The amplitude of gravitational waves is
\begin{equation}
  \label{eq:hmax}
  h = 5 \times 10^{-22} \left( \frac{r}{10\mbox{Mpc}} \right)
\end{equation}
and the total energy emitted is
\begin{equation}
  \label{eq:Etot}
  E_{\rscript{GW}}(t < 1 \mbox{msec}) = 10^{-3} \sol c^2
  = 2 \times 10^{51} \mbox{erg} \sim 0.05\% \mbox{ of }
  M_{\rscript{tot}} c^2 .
\end{equation}
The angular momentum of the emitted gravitational waves is given
by
\begin{equation}
  \label{eq:DelJ}
  \Delta J = \int \frac{dJ}{dt} \, dt ,
\end{equation}
\begin{equation}
  \label{eq:dJdt}
  \frac{dJ}{dt} = \frac{1}{16 \pi} \int \epsilon_{zjk} x^j
   ( \htt_{km,p} \TT{\dot{h}}_{mp} - \frac{1}{2} \htt_{mp,k}
   \TT{\dot{h}}_{mp} )  \, dS ,
\end{equation}
to be
\begin{equation}
  \label{eq:Jlost}
  \Delta J (t < 1 \mbox{msec}) = 1.8 \times 10^{-2} G \sol^2 / c \sim
  0.5 \mbox{\% of } J_t .
\end{equation}

\subsection{Irrotation case}

\begin{figure}[tbp]
  \begin{center}
    \leavevmode
    \epsfxsize=.7\textwidth \epsfbox{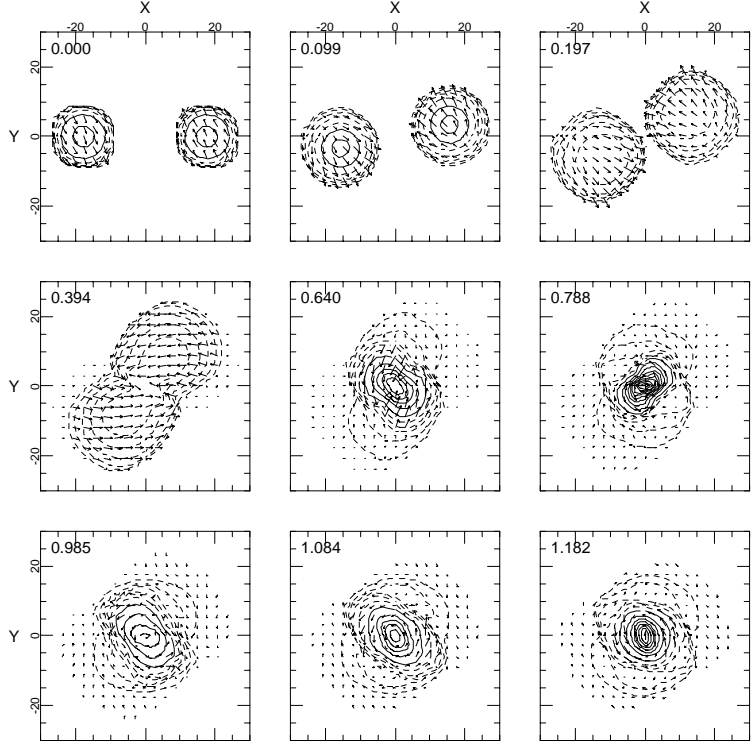}
    \caption{Density $\rhon$ and velocity $V^i$ on the $x$-$y$ plane
      for the irrotation case. Parameters are the same as
      Fig~\ref{fig:rho-cor}.}
    \label{fig:rho-irr}
  \end{center}
\end{figure}

The parameters of initial data for the irrotation case are shown in
the third column of Table~\ref{tab:initparm}. 
The evolution of the density and the velocity on the $x$-$y$ plane are 
shown in Fig.~\ref{fig:rho-irr}.\cite{WWW}

\begin{figure}[tbp]
  \begin{center}
    \leavevmode
    \epsfxsize=.75\textwidth \epsfbox{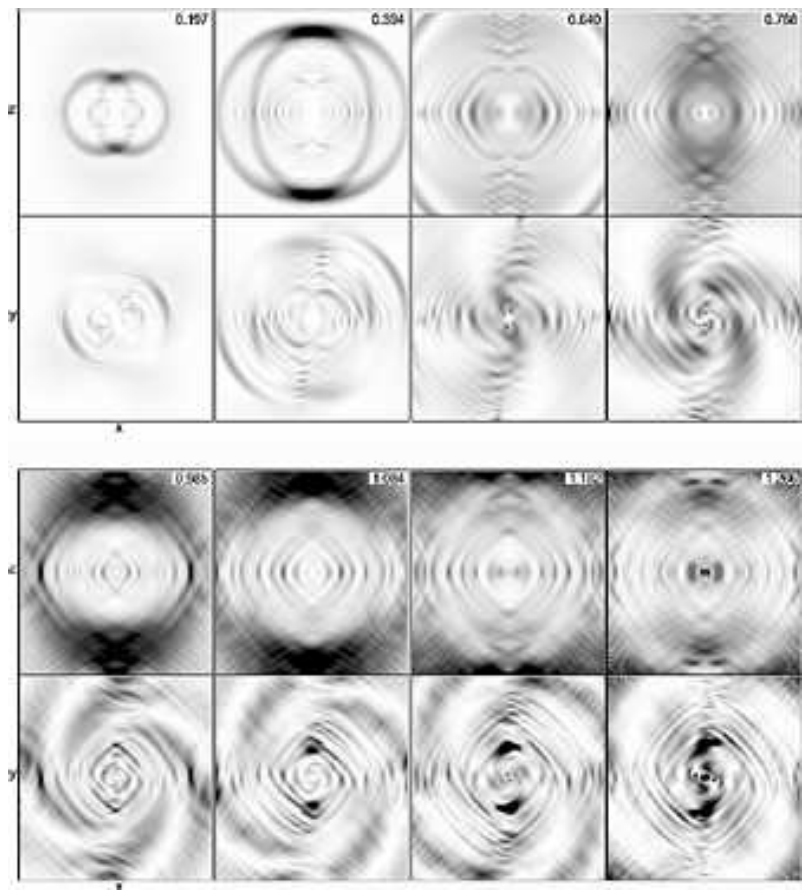}
    \caption{The propagation of the gravitational waves for the
      irrotation case. The ``energy densities of gravitational waves'' $r^2
      \rho_{_{\rscript{GW}}}$ on the $x$-$y$ and $x$-$z$ planes are
      shown similarly to Fig.~\ref{fig:gwxyxz-cor}}
    \label{fig:gwxyxz-irr}
  \end{center}
\end{figure}

\begin{figure}[tbp]
  \begin{minipage}[t]{.45\textwidth}
  \begin{center}
    \leavevmode
    \epsfxsize=\textwidth \epsfbox{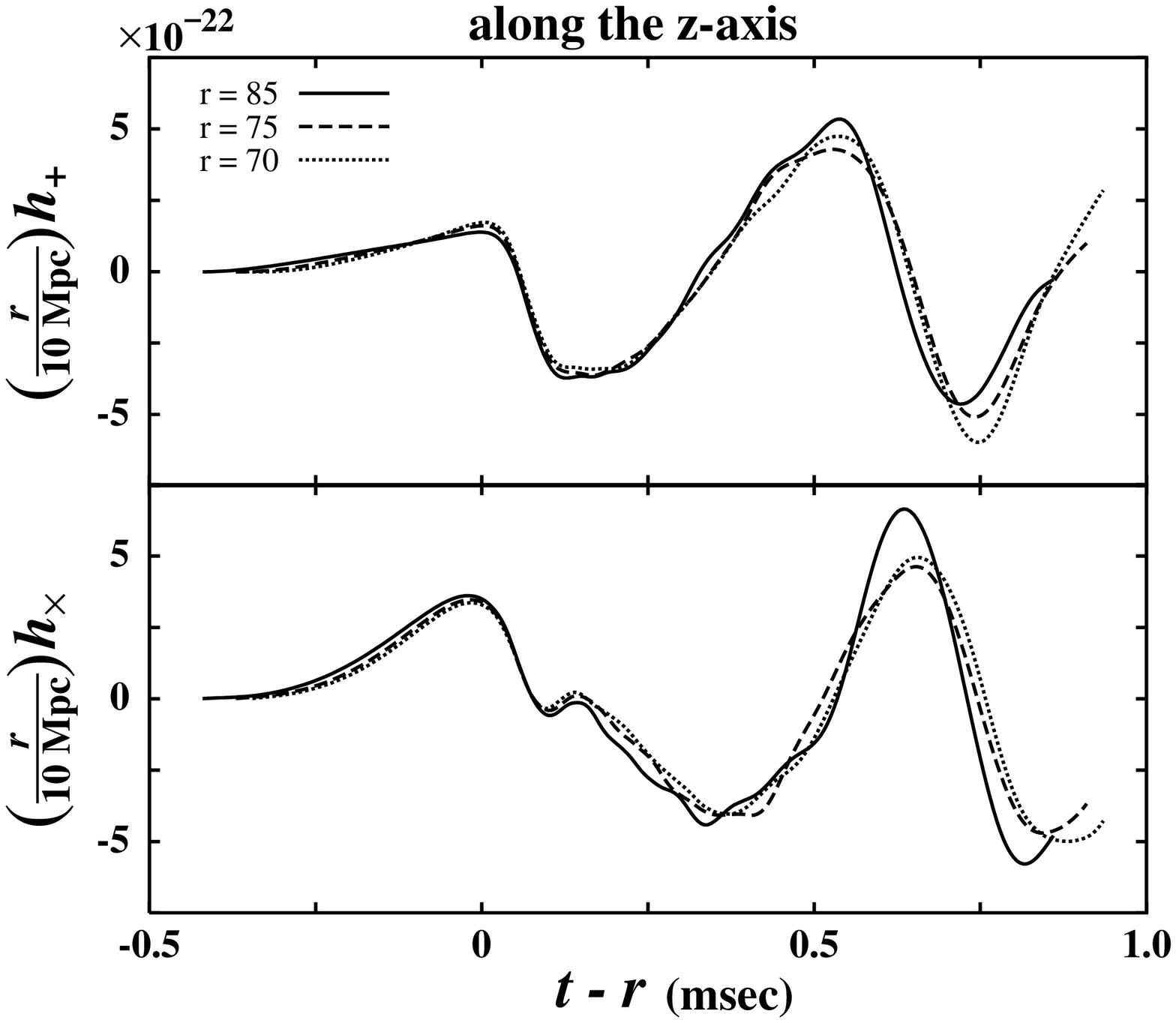}
    \caption{Waveforms $h_+$ and $h_\times$ observed on the $z$-axis
      for the irrotation case. They have been monitored at $r =
      75\sol, 80 \sol$ and $85 \sol$.}
    \label{fig:wave-irr}
  \end{center}
  \end{minipage}
  \hfill
  \begin{minipage}[t]{.5\textwidth}
  \begin{center}
    \leavevmode
    \epsfxsize=\textwidth \epsfbox{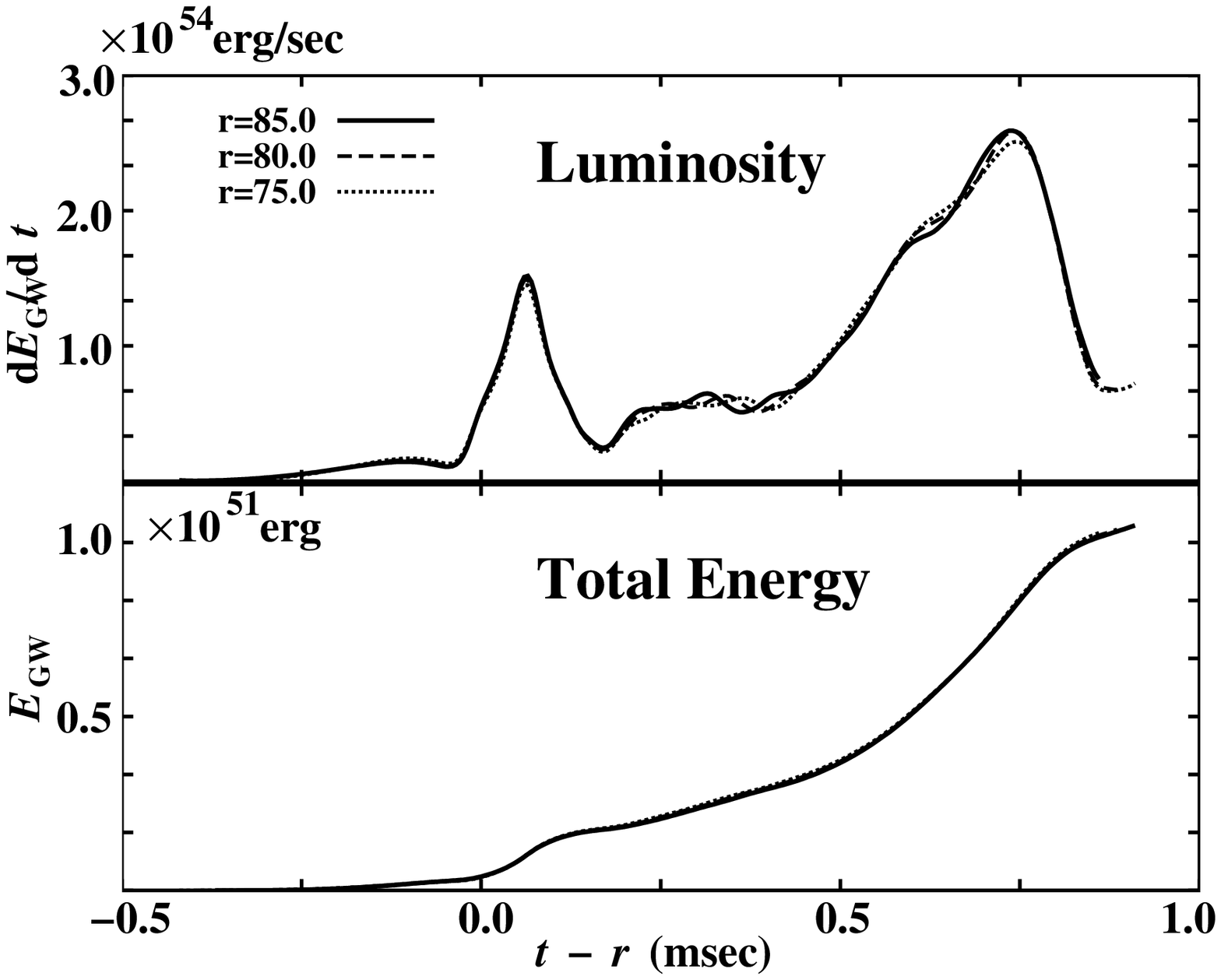}
    \caption{Luminosity (in units of erg/sec) and total energy (in
      units of erg) of the emitted gravitational waves.
      They are evaluated at $r = 75\sol, 80 \sol$ and $85 \sol$.}
    \label{fig:flux-irr}
  \end{center}
  \end{minipage}
\end{figure}

The propagation of the ``energy density of gravitational waves'' $r^2
\rho_{_{\rscript{GW}}}$ are shown in
Fig.~\ref{fig:gwxyxz-irr},\cite{WWW} the waveforms on the $z$-axis in
Fig.~\ref{fig:wave-irr} and the energy of the waves in
Fig.~\ref{fig:flux-irr}.

\begin{figure}[tbp]
  \begin{center}
    \leavevmode
    \epsfxsize=.45\textwidth \epsfbox{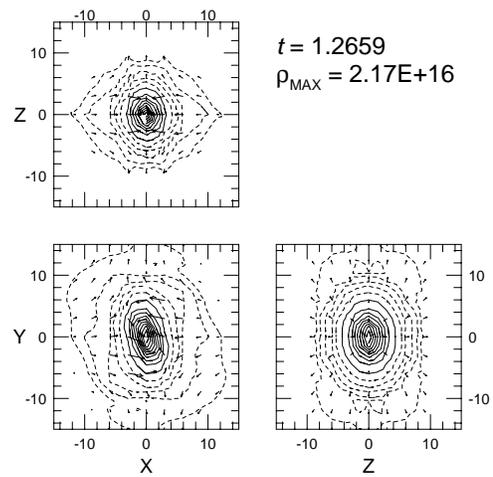}
    \caption{Density $\rhon$ and velocity $V^i$ in the central region
      on the $x$-$y$, $x$-$z$ and $z$-$y$ planes for the irrotation case.
      Parameters are the same as Fig.~\ref{fig:rhoc-cor}.}
    \label{fig:rhoc-irr}
  \end{center}
\end{figure}

The evolution is in principle the same as the corotation case, but no
outflow along the rotation axis appears in Fig.~\ref{fig:rhoc-irr}.
It is because the merger is progressing slower. It is the case in
post-Newtonian calculations also that the progress of the merger for
the irrotation case is slower than for the corotation case.\cite{ONS97}
It causes the emission of gravitational waves to decrease
somewhat. The amplitude, the energy and the angular momentum of the
gravitational waves emitted up to 1msec are, however, almost the same
as in the corotation case:
\begin{eqnarray}
  & & h = 5 \times 10^{-22} \left( \frac{r}{10\mbox{Mpc}} \right),
  \nonumber \\
  \label{eq:hEJirr}
  & & E_{\rscript{GW}}(t < 1 \mbox{msec}) \sim 0.05\% \mbox{ of }
  M_{\rscript{tot}} c^2, \\
  & & \Delta J (t < 1 \mbox{msec}) \sim 0.5 \mbox{\% of } J_t . \nonumber
\end{eqnarray}

\section{Summary}

We have described a numerical code for simulations of coalescing binary
neutron stars and presented some results with it. This code was
developed after extensive study of a post-Newtonian code for
coalescing binary neutron stars.\cite{ONS97} To solve hydrodynamics
equations and elliptic equations, the general relativistic
code uses the same methods  as the post-Newtonian code does.
For evolution equations for the metric, however, we adopted the
CIP method, which made the evolution of the metric stable even with a
coarse grid.

We faced instability when coalescence of two stars advanced to be
probably a black hole. It is caused by a characteristics of the
conformal slicing as well as by the coarseness of the grid. This fact may
suggest that other slicing such as the maximal slicing must be
adopted. Thus we are now making another code with the maximal slicing.
With the maximal slicing, we must solve a elliptic equation for
$\alpha$, which is obtained by setteing the right-hand side of
Eq.(\ref{eq:evoltrK}) equal to zero. The cost of solving this equation
is great. In addition, to catch the wave form of gravitational waves,
we have to apply a gauge-invariant wave extraction technique.\cite{GIWE}
Since new super computer, 7 times faster with 20 times greater memory,
will be available at KEK from the beginning of 2000, more precise
calculation with the maximal slicing will be carried out.
To calculate the right-hand side of Eq.(\ref{eq:beta}), we used the
extrapolated values for $\beta^i$ as describe in \S \ref{sec:ellip},
but the precision of the
extrapolation will reduce in the highly dynamical phase of
coalescence. We will be able to solve Eq.(\ref{eq:beta}) more
precisely by an iterative method on the new machine.

\section*{Acknowledgments}

The calculations presented in this paper were carried out on Fujitsu
VPP500/80 at High Energy Accelerator Research Organization as KEK
Supercomputer Project No.99-52 and on Fujitsu VPP300/16R at the National
Astronomical Observatory Japan.
In addition, this work was in part supported by the Grant-in-Aid of
Education, Culture, Science and Sports (09NP0801, 10640227).

\end{document}